\documentclass[11pt,a4paper]{article}
\usepackage{amsmath,amssymb} 
\usepackage{epsf} 
\begin{document}

\title{Nearby young single black holes}
\author{M.E. Prokhorov, S.B. Popov\\
{\it (Sternberg Astronomical Institute)}\\
mike@sai.msu.ru; polar@sai.msu.ru}
\date{}

\maketitle

We consider nearby young black holes formed after supernova explosions in
close binaries whose secondary components are currently observed as the
so-called runaway stars. Using data on runaway stars and making reasonable
assumptions about the mechanisms of supernova explosions and binary breakup,
we estimate the present position of nearby young black holes. For two
objects, we obtained relatively small error regions ($\sim 50$-100 deg$^2$).
The possibility of detecting these nearby young black holes is discussed.

\section{Introduction}

To date,\footnote{{\bf A note for the astro-ph version}. This paper was
published more than 3 years ago, however, it is not easily available via
Internet. So, we
submit it  now to the ArXiv
without any changes, including language editing, etc. (except several
clear mistakes), as it apeared in Astronomy Letters vol. 28, pp. 536-542
(2002). The original PDF file is also available at this URL:
http://xray.sai.msu.ru/$\sim$polar/html/sci.html.} stellar-mass black holes
(BHs) have been discovered in close binaries (see a review, for example, in
Cherepashchuk 1996) and supermassive BHs have been discovered in galactic
nuclei (for a review see Kormendy 2001). 
It would be of great interest to find a single
stellar-mass BH, but this is technically very difficult to do. Therefore,
nearby single BHs are of considerable interest. To detect such objects, it
would be desirable to reduce the search area, i.e., to estimate the
positions of possible sources in advance. Below, we suggest a method of such
estimation and use specific examples to illustrate it.

 Popov et al. (2002) briefly discussed nearby young compact objects (neutron
 stars and BHs) and proposed that radio-quiet neutron stars in the solar
 neihborhood were associated with recent supernova explosions that produced
 various structures in the local interstellar medium (Local Bubble, Loop I,
 etc.). Here, we analyze nearby young BHs in more detail.

The main idea of our study is as follows. We estimate the present positions
of nearby ($r<1$ kpc) young ($t<6$ Myr) BHs formed in close binaries with
massive secondary components that broke up after the first supernova
explosion. The so-called runaway star (Blaauw 1961) appears after binary
breakup. Knowing the present position and velocity of the runaway star and
specifying certain parameters for the binary and supernova explosion (see,
e.g. Lipunov et al. 1996 about the evolution of binary stars), we can
estimate the present-day position of a black hole.

\section{Young massive stars in the Solar neighborhood}

The galactic region where the Sun is located has some pecularities. The
so-called Gould Belt (P\"oppel 1997) dominates in the solar neighborhood.
This is a disk-like structure, $\sim$ 750 -1000 pc in size, whose center is
at 150-250 pc from the Sun. The plane of the Gould Belt in inclined
$\sim18^{\circ}$ with respect to the Galactic plane. The age of the Gould
Belt is estimated to be 30-70 Myr; i.e., the life of the most numerous stars
among those that can produce supernova explosions ($M\approx
8$-$10\,M_{\odot}$) is coming to an end there. Single radio-quiet neutron
stars discovered by the ROSAT satellite (see, for example, Popov et al.
2002) and some of unidentified EGRET sources (Grenier and Perrot 2001) are
probably associated with the Gould Belt.

Fifty-six runaway stars are known within $\sim$ 700 pc of the Sun
(Hoogerwerfet al. 2001). They are formed either during the dynamical
evolution of the clusters and associations where they were born (the most
likely cause is a close encounter of binaries) or through the binary breakup
during a supernova explosion. Four stars from this group have masses larger
than $\sim 30 M_{\odot}$ (since these stars are single and massive, the
accuracy of determining their masses is not very high).

Table 1 gives data [parameters from Hoogerwerf et al. 2001] on the runaway
stars considered here. Hoogerwerf et al. (2001) investigated all 56 nearby
runaway stars in detail. These are nearby stars in that they were studied by
HIPPARCOS satellite and their sky positions, proper motions, and parallaxes
are known within milliarcsecond accuracy (here, we ignore the errors in the
velocities and other parameters of the runaway stars). The authors traced
the motion of these stars in the Galaxy and for most of them (including the
four massive stars), they found when and from which association they escaped
and which of the two possibile ejection mechanisms operated for each
particular star.

\begin{table}[t]
\caption{Parameters of the four most massive runaway stars in the solar
neighborhood (Hoogerwerf et al. 2001)}
\begin{tabular}{l|c|c|c}
\hline
\multicolumn{1}{c|}{Star}& Mass, $M_\odot$ &
\parbox[c][1.3cm]{1.6cm}{Velocity,\\ km/s} &
\parbox[c][1.3cm]{3cm}{Kinematic \\ age, Myr}\\
\hline
  $\xi$ Per     & 33     & 65 & 1\phantom{.9} \\[3pt]
  HD 64760      & 25--35 & 31 & 6\phantom{.9}\\[3pt]
  $\zeta$ Pup   & 67     & 62 & 2\phantom{.9}\\[3pt]
  $\lambda$ Cep & 40--65  & 74 & 4.5\\
\hline
\end{tabular}
\end{table}

The four massive runaway stars are most likely to have acquired their high
space velocities through binary breakups after supernova explosions (to all
appearences, the fifth massive star, {\it i} Ori, was ejected from its
parent association through dynamical interaction; Hoogerwerf et al. 2001).
Several arguments may be advanced in support of this conclusion:

(1) These stars are very massive. To be ejected from the cluster
(association), they had to pass near stars of comparable mass. Otherwise,
according to the law of momentum conservation, less massive star would be
ejected from the cluster, whereas such massive stars are very few for any
reasonable mass function. Close encounters of several massive stars turn out
to be extremely rare events compared with rare close triple encounters of
low-mass stars.

(2) Massive stars live only several Myr. This imposes an additional
constraint on the rare events described above: the encounter must take place
until the massive star explodes as supernova.

(3)   Finally, all these stars move at velocities that are several times
higher than the velocity dispersion of their parent associations. This fact
does not contradict anything; after a succesful close encounter, the stars
can acquier high velocities. However, this occurs only in rare cases; the
mean velocity acquired in such processes is much lower.

More detailed arguments for each of the four stars from this group can be
found in Hoogerwerf et al. (2001).

 Thus, to all appearences, each of these four stars was a member of a binary
in which its neighbor exploded some time ago. The exploded star traversed
its entire evolutionary path faster; i.e., it was even more massive than the
observed runaway star. Such massive stars ($M>30$-$40\, M_{\odot}$) are
currently believed to collapse not into neutron stars but into BHs (White
and van Paradijs 1996; Fryer 1999). Moreover, the cores in stars with
slightly higher masses ($M\gtrsim 40 $-$50\, M_{\odot}$) are most likely to
collapse directly into BHs without going through the intermediate stage of a
hot neutron star (see, e.g. Bisnovatyi-Kogan 1968).

\section{Binary breakup after supernova explosion}

If a supernova explodes symmetrically in a binary with a circular orbit,
then {\it at least half of the binary mass} must be ejected for the binary
to break up [all aspects of binary breakup during mass ejection were
considered in detail by Hills (1983)]. For example, if the mass of the
runaway star $M_{opt}=30\,M_{\odot}$ and if it did not change significantly
since the binary breakup, while the BH mass is $M_{BH}=10\,M_{\odot}$, then
the mass of the ejected envelope must be no less than $\Delta M \geq
M_{opt}+M_{BH}=40\, M_{\odot}$ and the mass of the exploded presupernova is
$M_{SN}=M_{BH}+\Delta M \geq 50\, M_{\odot}$. Since the mass loss from such
massive stars over their lifetime is large (at least 30\% of the initial
mass), each of the stars under consideration was a member of an {\it
extremely massive} binary. The presupernova mass for $\zeta$ Pup that
follows from such reasoning is 87.5~$M_{\odot}$; i.e., either this was a
particularly massive star ($>100\, M_{\odot}$ during its birth) or the mass
loss was much lower than that predicted.

We consider only binaries with two massive stars and assume that none of
their components filled their Roche lobe before a supernova explosion. Note
that it is unlikely that these systems passed through the stage of mass
transfer. However, if such a process takes place, then for both stable and
unstable (with a common envelope) mass transfer, the primary component will
lose part of its mass and the mass of the secondary component will be
constant or increase. As a result, the binary-component mass ratio decreases
and a symmetric supernova explosion will most likely be unable to tear the
binary apart. Our second condition, a circular orbit, is guaranteed to be
satisfied after the stage of mass transfer.

Since the binaries under consideration are close systems (the current
velocities of the runaway stars are on the order of their orbital velocities
in binaries), the assumption of circular orbits appears acceptable and the
high presupernova mass makes probable the direct collapse of the supernova
core into a BH (White and van Paradijs 1996). Such collapse is generally
believed to be symmetric and without recoil (i.e., the BH velocity is the
same as that of the presupernova velocity before the explosion). This is in
contrast to the formation of neutron stars, which are born with space
velocities of several hundred kilometers per second (Lyne and Lorimer 1994).

Binary breakups through supernova explosions were considered by several
authors (see, e.g., Tauris and Takens 1998; Hills 1983). However, since the
above two conditions are most likely satisfied, the breakup proceeds in a
simple way (see Fig. 1). The envelope is ejected symmetrically about the
presupernova center and is carried away in a straight line in the direction
and with the velocity of its orbital motion at the explosion time. The
motion referes to the center of the envelope and is unaffected by its
symmetric expansion. The center of mass of the two stars (the BH and the
binary's secondary component, which became a runaway star) moves in the
opposite direction but at higher velocity, because the mass of the ejected
envelope exceeds the total mass of the remaining stars.

\begin{figure}
\epsfxsize=0.7\textwidth
\centerline{\epsfbox[213 353 405 520]{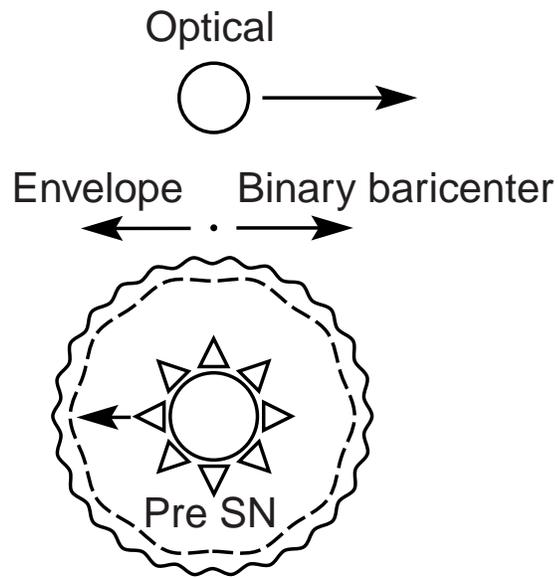}}
\caption{A scheme for  binary breakup after a supernova
explosion.}
\end{figure}

\begin{figure}
\epsfxsize=0.5\textwidth
\centerline{\epsfbox{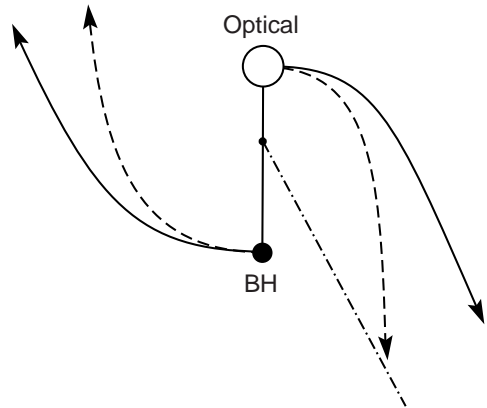}}
\caption{Star separation  after a supernova explosion in the presupernova
center-of-mass frame of reference.}
\end{figure}

   In the center-of-mass frame of reference of the two stars (without the
ejected envelope), the star velocities immediately after an explosion are
directed perpendicularly to the line that connects them and the realtive
velocity of the star and the BH is equal to the relative orbital velocity of
the stars before the explosion (see Fig.~2). The runaway star and the BH
move along similar hyperbolas with the eccentricity $e=\Delta
M/(M_{opt}+M_{BH})\geq $1. As the two stars move apart, the vectors of their
velocities turn through angle $\varphi: {\mathrm{sin}} \varphi =1/e$. In the
limiting case where the ejected mass is exactly equal to half the binary
mass, the stars move along parabolas ($e=1$) and the derection of their
velocities change by $90^{\circ}$ in the separation time. The parabolic
trajectories are indicated by the dashed lines in Fig.~2.

In the presupernova center-of-mass system (Fig~3), the hyperbolic or
parabolic star separation is supplemented with the uniform motion of their
center of mass. As a result, both the runaway star and the BH move in the
direction opposite to the motion of the ejected envelope.

\begin{table}[t] 
\caption{Parameters of the error regions for the BHs associated with massive runaway stars}
\begin{tabular}{l|c|c|c|c}
\hline
\multicolumn{1}{c|}{Name}& Distance, pc  &
 Velocity, km/s & Localization area & $N_{\textrm{EGRET}}$\\
\hline
  $\xi$ Per     &  537--611 & 19--70 & $\sim  7^{\circ} \times  7^{\circ}$
&\phantom{9}1  \\[3pt]
  HD 64760      &  263--645 & 11--59 & $\sim 45^{\circ} \times 50^{\circ}$
&12 \\[3pt]
  $\zeta$ Pup   &  404--519 & 33--58 & $\sim 12^{\circ} \times 12^{\circ}$
&\phantom{9}1  \\[3pt]
  $\lambda$ Cep &  223--534 & 19--70 & $\sim 45^{\circ} \times 45^{\circ}$
&\phantom{9}6  \\
\hline
\end{tabular}
\end{table}

\section{Calculating the BH position}

The errors in the proper motions and parallaxes of stars affect the {\it
relative} positions of the BH and the runaway stars only slightly. The
contribution of these errors to the BH localization is less significant than
the uncertainties in the remaining parameters. Given the sky position of
each of the stars, their distance, and velocity component, we can integrate
the star's motion in the Galactic gravitational field back in time. We took
the kinematic age (the time elapsed since the supernova explosion and binary
breakup) from Hoogerwerf et al. (2001). Therefore, we can determine the
relative velocity $v_{opt}$ with which each of the runaway stars escaped
from its parent association and its direction. The BH velocity $v_{BH}$ must
be determined from $v_{opt}$. The problem has a unique solution if $\Delta
M$ and $M_{BH}$ are known, and we can find the velocity $v_{BH}$ and the
angle $\psi$ that it makes with $v_{opt}$:
$\psi(v_{{BH}},v_{{opt}})=
\widehat{{\mathbf{v}}_{{BH}}{\mathbf{v}}_{{opt}}}$. The center of mass of
the envelope, the BH, and the runaway star moves in the orbital plane of the
binary whose orientation is unknown. Thus, the velocity
${\mathbf{v}}_{{BH}}$ is directed along the side surface of the cone whose
axis coincides with ${\mathbf{v}}_{{opt}}$ and whose half-angle is $\psi$.
We characterize the specific position of vector ${\mathbf{v}}_{{BH}}$ on the
cone by an azimuthal angle $\phi$ ($\phi$ is related to the orientation of
the binary orbital plane; the choice of the zero point from which it is
counted off is of no importance for subsequent analysis). Since we cannot
determine the specific position of ${\mathbf{v}}_{{BH}}$ on the cone surface
(i.e., $\phi$) from observations, this parameters must be varied.

\begin{figure}
\epsfxsize=0.5\textwidth
\centerline{\epsfbox{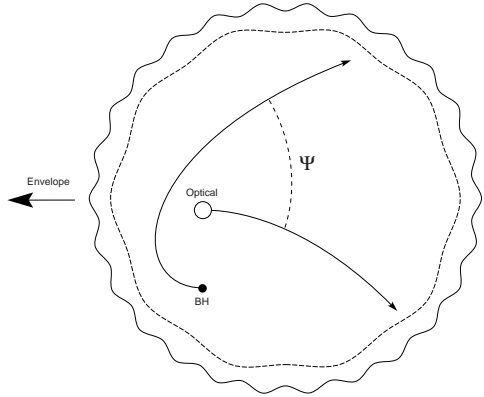}}
\caption{Binary flying apart after the supernova explosion as seen
in the presupernova center-of-mass frame.}
\end{figure}

\begin{figure}
\epsfxsize=0.8\textwidth
\centerline{\epsfbox{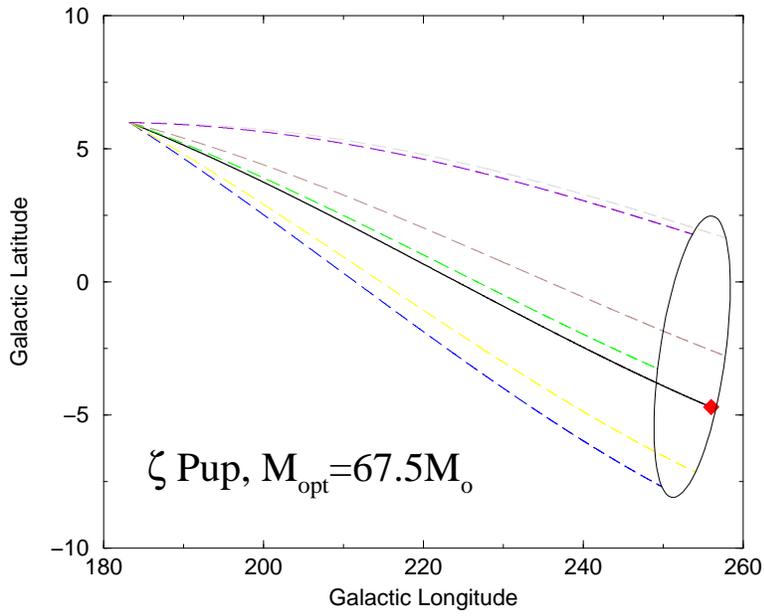}}
\caption{Sky trajectory of the runaway star $\zeta$~Pup  (solid line)
and four possible trajectories
of the black hole (dashed lines). We assumed that the mass of the black hole is
$M_{{BH}}={10{M}_\odot}$. }
\end{figure}

 After specifying the binary breakup parameters, we must integrate the
motion of the BH from its birth and to the present time. To integrate 
the motion in the Galactic potential, we used the same code and constants
specifying the Galactic potential as in our previous computations of the motion of single neutron stars (Popov et al. 2000).

  Here we make three simplifying assumptions, which are discussed below:

\begin{itemize}
\item the supernova explosion is symmetric, i.e., the space velocity of the remnant (BH) does not vary during the explosion; 
\item the association moves in a circular orbit in the Galactic disk;
\item the binary velocity inside the association is disregarded.
\end{itemize}

These assumptions allow us to use the above relation between the velocities
of the runaway star and BH at the point of binary breakup. For each set of
parameters $\phi$, $\Delta M$, and $M_{BH}$, we obtain the vector 
$\mathbf{v}_{BH}(\phi, \, \Delta M, \, M_{BH})$. Integrating the BH motion 
from the supernova explosion to the present time, we find its sky position. 
When exhausting the admissible values of the parameters, these points
 sweep the sky region where the BH must be searched for.

\begin{figure}
\epsfxsize=0.8\textwidth
\centerline{\epsfbox{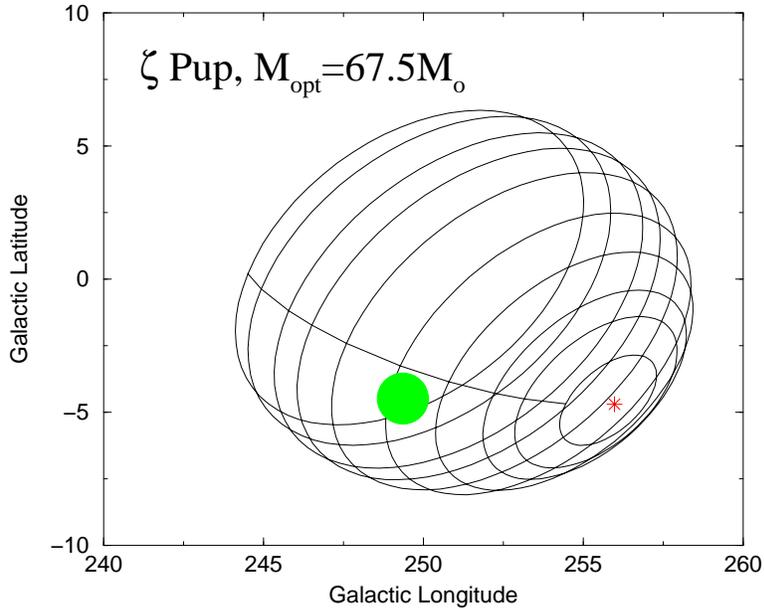}}
\caption{The possible localization area for the black hole that
originated from the same disrupted binary as the runaway star $\zeta$~Pup.
The rings correspond to different ejected
masses $\Delta M$ and orientations $\phi$ of the presupernova orbit.
The asterisk and the circle indicate the positions of the runaway star
and that of the unidentified EGRET source (3EG~J0747--3412).
We assumed that the mass of the black hole is
$M_{{BH}}=10{M}_\odot$. The smallest $\Delta M$ corresponds
to the ring that is nearest
to the runaway star. }
\end{figure}

\begin{figure}
\epsfxsize=0.8\textwidth
\centerline{\epsfbox{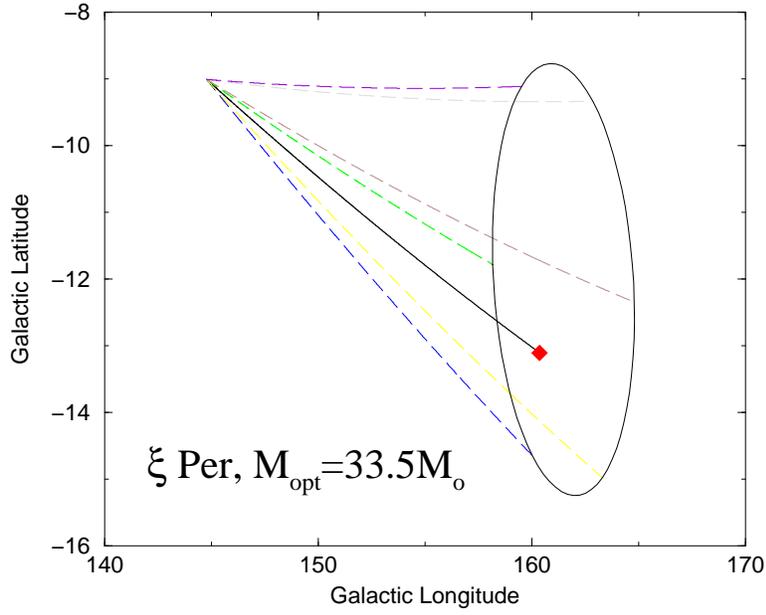}}
\caption{Same as Fig.~4 but for the runaway star $\xi$~Per.}
\end{figure}

\begin{figure}
\epsfxsize=0.8\textwidth
\centerline{\epsfbox{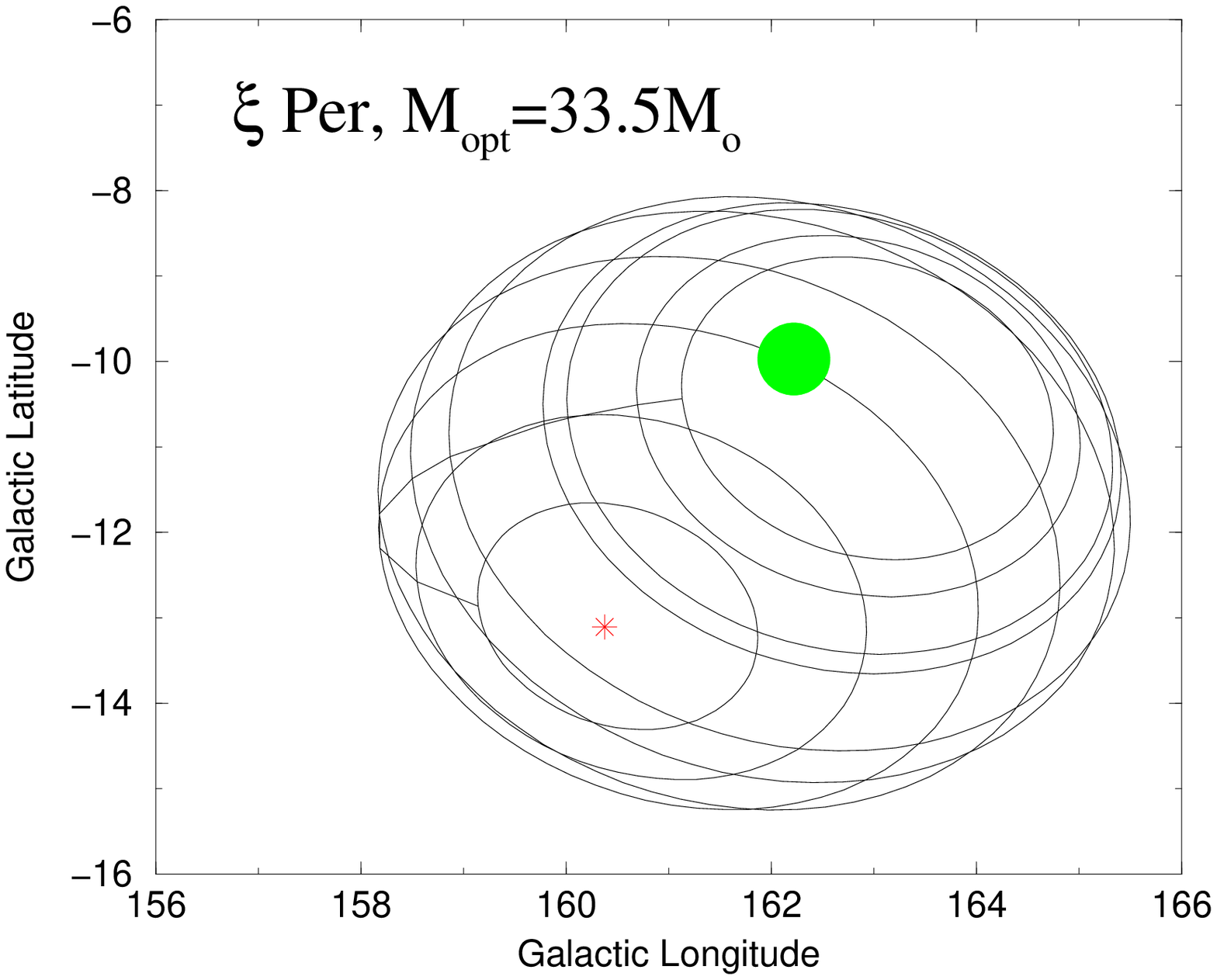}}
\caption{Same as Fig.~5 but for the runaway star $\xi$~Per.
The circle indicates the position of
the unidentified EGRET source (3EG~J0416+3650). }
\end{figure}

Table 2 gives  the following data on BHs: the heliocentric distance; the BH velocity relative to the interstellar medium (i.e., relative to the circular Galactic rotation at a given point); the size of the error region; and the number of unidentified EGRET sources in this region. Despite the simplifying assumptions,
we obtained large regions in the sky for $\lambda$ Cep and HD64760 in which the search was not promising. Figure 4 shows the trajectory of the optical star 
and a number of possible BH trajectories  for the binary that produced $\zeta$ Pup. Figure 5 shows the possible BH error region for this system (both figures are in Galactic coordinates). Figures 6 and 7 show the same results for $\xi$ Per. Since we obtained large error regions for other stars, no similar figures are given here for them.

The mass of the ejected envelope $\Delta M$, the presupernova mass $M_{SN}$,
and the BH velocity relative to the interstellar medium for the rings are shown below for $\xi$ Per and $\zeta$ Pup.

{\noindent\small
\begin{tabular}{lcccccccccc}
\\
{\normalsize$\mathbf\zeta$~{\bf Pup}}\\
$\Delta M, M_\odot$ & 78 & 79 & 80 & 82 & 85 & 90 & 95 & 100 & 110 & 120 \\
$M_{SN},   M_\odot$ & 88 & 89 & 90 & 92 & 95 & 100& 105 & 110 & 120& 130  \\ 
$v$, km/s& 57--58 & 56--57 & 55--56 & 53--55 & 51--52 &
47--49 & 44--46 & 41--43 & 37--38 & 33-35 \\
\\
{\normalsize$\mathbf\xi$~{\bf Per}}\\
$\Delta M, M_\odot$ & 44 & 45 & 47 & 50 & 55 & 60 & 75 & 80 & 100 & 120  \\
$M_{SN},   M_\odot$ & 54 & 55 & 57 & 60 & 65 & 70 & 85 & 90 & 110 & 130\\
$v$, km/s & 69--70 & 66--68 & 62--63 & 56--58 & 49--51 & 44--46 &
33--35 & 31--32 & 24--25 & 19-20\\
\\
\end{tabular}

}

 The largest masses are given for illustrative purposes. It should be noted,
 however, that a reduction in the upper limit of the presupernova mass to
 100 $M_{\odot}$ for $\xi$ Per and to 120 $M_{\odot}$ for $\zeta $Pup
 changes the BH error regions only slightly.

\section{Discussion and conclusions}

The errors in the proper motions and parallaxes affect only slightly the
relative positions of the BHs and runaway stars. The contribution of these
errors to the BH localization error is less significant than that of the
uncertainty in other parameters ($\Delta M, \, M_{BH}, \, \phi$). From our
assumptions about the velocities, the first assumption (about zero recoil
during the BH formation) seems most uncertain. If we draw an analogy with
neutron stars (i.e., if we scale the velocity in accordance with an increase
in the mass of the compact object and with changes in other parameters),
then the BH could gain an additional velocity of up to several tens km/s
during its birth, which would completely change the inferred error region.
However, as yet, no compelling experimental evidence is available for a low
or high BH recoil velocity.

 The assumption about the circular motion of young stellar associations in
the Galactic disk appears plausible enough. Moreover, this motion can, in
principle, be measured. The motion of a binary inside an association can be
taken into account in calculations (by adding a randomly oriented velocity
on the order of the velocity dispersion inside the association to the
velocity vector of its center). These velocities are small and taking them
into account increases only slightly the size of the localization areas (by
15--20 $\%$ according to our estimates). We ignore these small corrections
in this paper.

The probability of finding a black hole increases towards the corresponding
runaway star. This is due to the fact that closer sky positions of the two
components corresponds to closer pre-explosion masses. Small separations
between the black hole and the runaway star prove to be the most probable
for mass functions that fall off toward large masses. However, the
distribution function is wide enough to show no sharp maximum at the
present-day position of the runaway star. The situation is further
complicated if seen from observational viewpoint, because greater separation
between the black hole and the runaway star corresponds to lower space
velocity of the former, i.e., to higher accretion rate at the same
interstellar-gas density, thereby making detection of such an object easier.

The activity of the black hole at hard wavelengths may be associated with
the accretion of turbulized interstellar matter (Shvartsman 1971). Such
matter has nonzero angular momentum preventing it from
falling immediately onto the black hole. Instead it forms an accretion ring
in the vicinity of the latter, which, due to viscosity, transforms into an
accretion disk. If the matter from such the ring does not
fall completely onto the black hole during the time it takes it to cross
an interstellar turbulence cell, another ring with a different orientation
begins to form near the black hole. These rings "annihilate" (i.e., mutually
destroy each other), thereby increasing the accretion rate, which varies strongly over time
scales on the order of the turbulent cell crossing times (from a few days to
several years depending on the velocity of the black hole relative to the
interstellar medium). The upper limits on the black hole velocity (see Table~2)
are rather high and no important accretion rate is to be expected in this case,
however, the lower limits make the $\dot M$ estimate rather optimistic.

Many authors analyzed the efficiency of accretion onto a single black hole
(see, e.g., Gruzinov (1998) and references therein). However, of special
interest in our case is nonstationary activity of black holes (Gruzinov
1999). In the case of low mean luminosity and relatively large heliocentric
distances (hundreds of pc, see Table~2) the source can be unscreened during
the short-term flux increase.

For stars $\xi $~Per and $\zeta$~Pup we obtained relatively small
possible black-hole localization areas and therefore for each of these objects
only one candidate can be found in the
third EGRET catalog. These sources are 3EG~J0747--3412
(for $\zeta$~Pup) and 3EG~J0416$+$3650 (for $\xi $~Per). For $\lambda$~Cep
and HD~64760, for which our computations yielded large black-hole
localization areas, we found a total of 6 and 12 sources, respectively, in
the EGRET catalog. However, in the latter case of special interest may be
the sources that were observed especially close to the runaway star. These are
3EG~J2227$+$6122 in the case
of $\lambda$~ Cep and 3EG~J0724--4713, 3EG~J0725--5140, 3EG~J0828--4954,
and 3EG~J0903--3531 in the case of HD~64760.

Note that an analysis of massive runaway stars may shed additional light
onto the explosion mechanisms of massive stars. According to the now most
popular supernova explosion mechanism (Fryer 1999), the collapse of a star
of mass $> 40 M_{\odot}$  proceeds with no mass ejection at all and ends up
with the formation of the most massive black holes. However, in this
case it is difficult to explain the disruption of binaries with the second
components having  masses greater than $\sim 30 M_{\odot}$. Prokhorov and
Postnov (2001) analyzed various supernova explosion mechanisms and concluded
that the observed distribution of compact objects agrees best with what is
obtained as a result of magnetorotational mechanism. This mechanism
is characterized by much weaker recoil of black holes compared to that of
neutron stars and, moreover, the envelope
is ejected even if a black hole forms. A study of the products
of the disruption of close binary systems may provide additional
arguments in favor of
some supernova explosion mechanism.

Besides black holes that formed in massive close binary systems
the solar neighborhood must also contain about 20 black holes younger than
10~Myr. This follows from the supernova rate in the Gould Belt,
which is about 20--30 per Myr (Grenier 2000) and the ratio of the
number of neutron stars to that of black holes (on the order of $10:1$).
Moreover, we can expect a large number of old black holes to exist
within 1~kpc from the Sun. These objects are, however,
difficult to identify without some a priori knowledge about their
coordinates and other parameters (space velocity, heliocentric distance).
That is why we tried to show how these parameters can be determined from
the data on runaway stars.

\section*{Acknowledgments}

We are grateful to Profs. Kuimov and Rastorguev for their advice and to the
reviewers for their comments that contributed to the improvement of this paper.
S.B.Popov thanks Monica Colpi, Aldo Treves, Roberto Turolla, and Luca Zampieri.
We are also grateful to the organizers
and participants of HEA-2001 conference (Space Research Institute of the
Russian Academy of Sciences) for providing a forum for the presentation
and fruitful discussion of this work. This work was supported by the Russian
Foundation for Basic Research
(grant no.~00-02-1716).

{\it Translated from russian by A. Dambis}

\end{document}